# Ultrasound cavitation and exfoliation dynamics of 2D materials revealed in operando by X-ray free electron laser megahertz imaging

Kang Xiang [1], Shi Huang [1], Hongyuan Song [1], Vasilii Bazhenov [2], Valerio Bellucci [2], Sarlota Birnšteinová [2], Raphael de Wijn [2], Jayanath C. P. Koliyadu [2], Faisal H. M. Koua [2], Adam Round [2], Ekaterina Round [2], Abhisakh Sarma [2], Tokushi Sato [2], Marcin Sikorski [2], Yuhe Zhang [3], Eleni Myrto Asimakopoulou [3], Pablo Villanueva-Perez [3], Kyriakos Porfyrakis [4], Iakovos Tzanakis [5], Dmitry G. Eskin [6], Nicole Grobert [7], Adrian Mancuso [2, 8, 9], Richard Bean [2], Patrik Vagovič [2, 10], Jiawei Mi [1,*]

[1] School of Engineering, University of Hull, Hull HU6 7RX, UK
[2] European XFEL, Holzkoppel 4, 22869 Schenefeld, Germany
[3] Synchrotron Radiation Research and NanoLund, Lund University, Box 118, 221 00, Lund, Sweden
[4] Faculty of Engineering and Science, University of Greenwich, Central Avenue, Chatham Maritime, Kent, ME4 4TB, UK
[5] School of Engineering, Computing and Mathematics, Oxford Brookes University, College Cl, Wheatley, Oxford, OX33 1HX, UK
[6] Brunel Centre for Advanced Solidification Technology, Brunel University London, Kingston Lane, London, UB8 3PH, UK
[7] Department of Materials, University of Oxford, Parks Road, Oxford OX1 3PH, UK
[8] Department of Chemistry and Physics, La Trobe Institute for Molecular Science, La Trobe University, Melbourne, Victoria, Australia
[9] The Diamond Light Source Ltd, Harwell Science & Innovation Campus, Didcot, OX11 0DE, UK
[10] Centre for Free-Electron Laser Science CFEL, Deutsches Elektronen-Synchrotron DESY, Notkestr. 85, Hamburg, 22607, Germany

* Corresponding author: j.mi@hull.ac.uk

## Abstract

Ultrasonic liquid phase exfoliation is a promising method for the production of two-dimensional (2D) layered materials. A large number of studies have been made in investigating the underlying ultrasound exfoliation mechanisms. However, due to the experimental challenges for capturing the highly transient and dynamic phenomena in real-time at sub-µs time and µm length scales simultaneously, most theories reported to date still remain elusive. Here, using the ultra-short X-ray Free Electron Laser pulses (~25ps) with a unique pulse train structure, we applied MHz X-ray Microscopy and machine-learning technique to reveal unambiguously the full cycles of the ultrasound cavitation and graphite layer exfoliation dynamics with sub-µs and µm resolution. Cyclic fatigue shock wave impacts produced by ultrasound cloud implosion were identified as the dominant mechanism to deflect and exfoliate graphite layers mechanically. For the graphite flakes, exfoliation rate as high as ~5 Å per shock wave impact was observed. For the HOPG graphite, the highest exfoliation rate was ~0.15 Å per impact. These new findings are scientifically and technologically important for developing industrial upscaling strategies for ultrasonic exfoliation of 2D materials.

## 1. Introduction

Two-dimensional (2D) functional materials have been intensively studied recently because of their unique properties and enormous application potentials in energy storage, catalysis, and nanoelectronics [1, 2, 3, 4]. The current challenge of manufacturing 2D materials is to develop efficient, cost-effective, and sustainable technologies for making large 2D sheets in substantial quantities [5, 6, 7, 8, 9]. Ultrasonic liquid phase exfoliation (ULPE) has been identified as one of the most promising technical routes for producing 2D large-size monolayer flakes (typically ~1 µm lateral size and ~1-10 atomic layers thick) with far fewer defects and less surface oxidation [10, 11, 12]. In the ULPE process, there are complex and dynamic interactions among the liquid, ultrasonic pressure wave, cavitation bubbles and solid materials across multi-length and multi-time scale [13, 14, 15]. Extensive research has been carried out in the past decade to understand the dynamic links between bubble dynamics, shock wave generation/transmission and layer exfoliation [13, 14, 16, 17, 18]. For example, Li *et al*. [19] suggested that the ultrasonic bubbles produced in sonication broke the graphene sheets into nanofragments. Cravotto *et al.* [5] and Alaferdov *et al.* [20] pointed out that cavitation-induced shock waves exerted high-density elastic acoustic energy onto the sheets to facilitate the polycrystalline graphite exfoliation in liquid phase. Lotya *et al.* [21] reported that the shock waves produced at the collapse of bubbles could act on the crystal edges, overcoming the Van Der Waals attraction between adjacent layers. However, these arguments were all based on the post-mortem microstructural analyses without *in-situ* observed evidence. Our recent *in-situ* and real-time studies using ultrafast synchrotron X-ray [14, 22] and high-speed optical imaging [13, 23] indicated that graphite layer exfoliation under ultrasound is a far more complicated process than previously proposed. The onset and growth of the exfoliated layers often occurred in a cyclic fatigue manner closely linked to the amplitude and time of the ultrasonic waves applied [14]. However, due to the limited contrast at the required megahertz temporal resolutions of synchrotron X-ray imaging, the sub-µs scale layer exfoliation dynamics at the instant of bubble implosion and the role of shock wave have not been fully quantified.

Here, we present our very recent work on using MHz XFEL Microscopy [24] (1.13M fps, a temporal resolution of 886 ns, and a pixel size of 3.2 µm) to study in operando the highly transient phenomena of ultrasonic cavitation dynamics, bubble implosion, shock wave emission and their effects on the onset of layer exfoliation and subsequent layer growth dynamics of bulk graphite materials. The MHz full field imaging capability provides incontrovertible evidence of the absolute dominance of the shock waves generated from the cavitation bubble collapse on the exfoliation of 2D materials. These findings are essential for deep understanding of the layer exfoliation dynamics in different ultrasound conditions and are critical for the optimization of the ULPE of 2D functional materials.

## 2. Methods

The experiments were carried out at the SPB/SFX instrument of the European X-ray Free-Electron Laser (EuXFEL) in Schenefeld, Germany [25, 26]. Fig. 1 shows the X-ray pulse trains and samples.

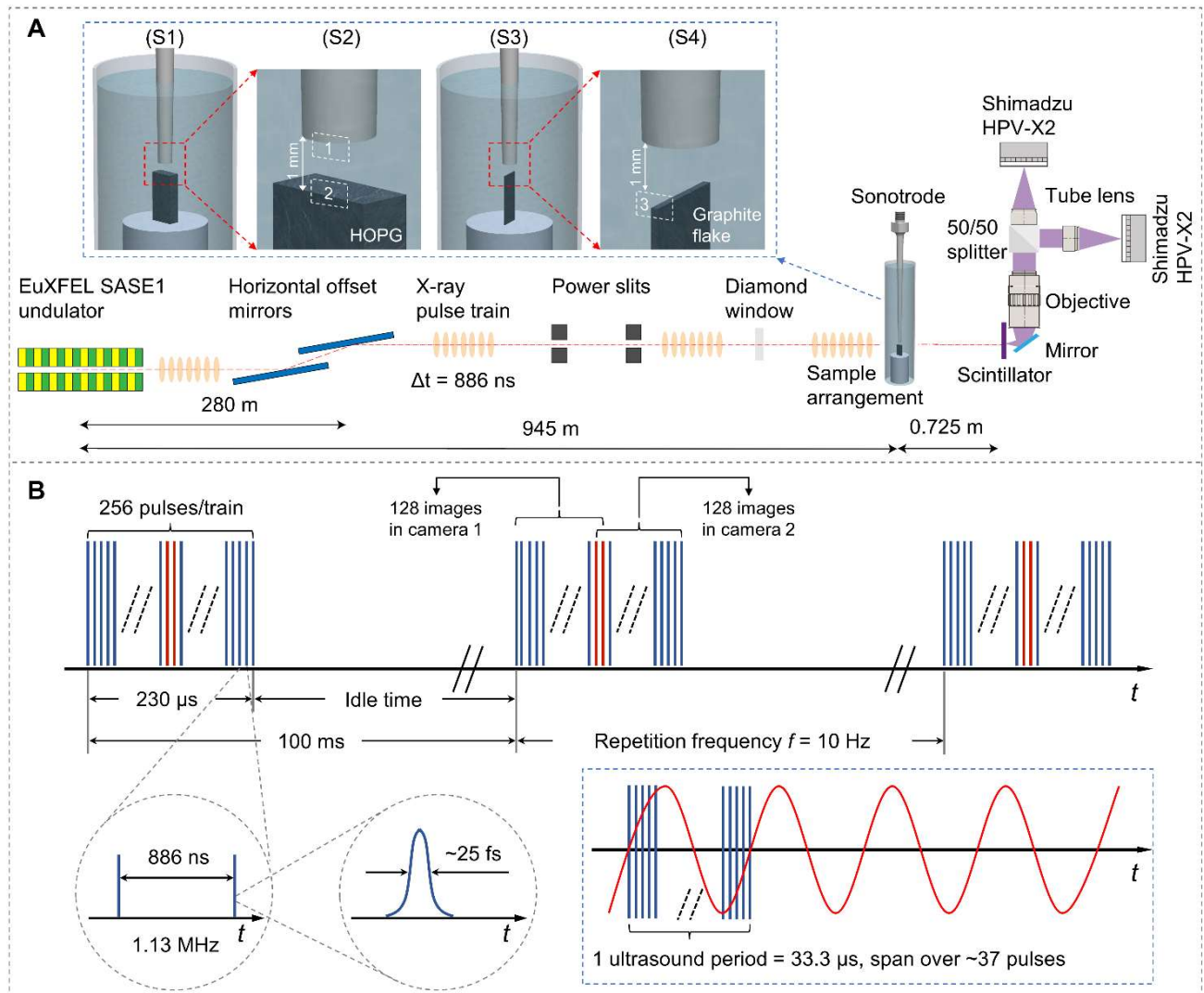

*Fig. 1. (A) The X-ray pulse trains (20 keV), sample arrangement and 2 Shimadzu HPV-X2 full-field cameras. (S1) and (S3) show the quartz tube sample holder, containing the deionised water, sonotrode and graphite samples. The white rectangles 1, 2, and 3 in (S2) and (S4) show the locations for imaging ultrasonic cavitation bubbles and interactions with the graphite sheet and flake. (B) The X-ray pulse train time structure. 37 pulses in 1 ultrasound period (i.e., 37 images in 1 ultrasound period with ~25 fs exposure time each).*

The highly oriented pyrolytic graphite sheets (HOPG, 10×3.5×1.5 mm from Agar Scientific Ltd) and the graphite flakes (GF, 2-3 mm with 0.2-0.4 mm thickness from Sigma-Aldrich) were used in the experiments (see Fig.1A). They were glued on the top of a specially designed stainless-steel sample holder by using an Sn-43%Bi solder alloy. A quartz tube (20 mm outer diameter, 80 mm tall, and 0.5 mm thick) with an inner volume of ~18 ml was used to hold ~15 ml deionized water (DiWater). A Hielscher UP100H ultrasonic processor with a fixed frequency of 30 kHz and a Ti-6Al-4V alloy sonotrode (MS2, 2.0 mm diameter tip, acoustic power intensity of 207 W/cm$^2$) were used to produce and transmit ultrasonic waves into the DiWater. 3 different locations (marked by the white rectangle 1, 2 in S2, and 3 in S4 of Fig. 1A) were selected to image the cavitation bubble dynamics, shock wave emission, and interactions with the graphite materials. Synchronization among triggering of the ultrasonic processor, image recording by the two Shimadzu cameras, and X-ray pulses was realized by using a synchronized microTCA timing board and Stanford DG645 digital delay/pulse generators for setting the time delay for each event.

Fig. 1B shows the X-ray pulse train time structure [27]. The train duration used was 230 μs (600 μs maximum) with 256 pulses in 1 train (~7 ultrasound cycles). Each pulse has an energy of ~700 μJ and a duration of ~25 fs. The time interval between two consecutive pulses within the train is 886 ns, providing an imaging frame rate of ~1.13 MHz. The repetition frequency of the pulse train is 10 Hz. A YAG: Ce 20 μm scintillator was used and the sample-to-scintillator distance was set at ~0.725 m. Two Shimadzu HPV-X2 full-field cameras coupled with a 10× NUV objective lens were used together to capture 254 images in one pulse train (the 127$^{th}$ and 128$^{th}$ images in the 1$^{st}$ camera overlaps with the 1$^{st}$ and 2$^{nd}$ images in the 2$^{nd}$ camera). 15 s is needed for transferring the image data until the next set of images can be acquired. The field of view (FOV) is 1280 (horizontal) × 800 (vertical) μm$^2$ with a pixel size of 3.2 μm. The collected images were flat-field corrected on a shot-to-shot basis using principal component analysis [28] and machine learning with online processing implemented [29, 30].

**3. Results & Discussion**

*3.1 Nucleation, growth and implosion dynamics of ultrasonic cavitation bubbles*

To study the nucleation and evolution of ultrasonic cavitation bubbles in different ultrasound intensities, we imaged systematically the bubble behaviour produced at different peak-to-peak amplitudes (just use amplitude hereafter) of the sonotrode: 9 μm, 20 μm, and 96 μm.

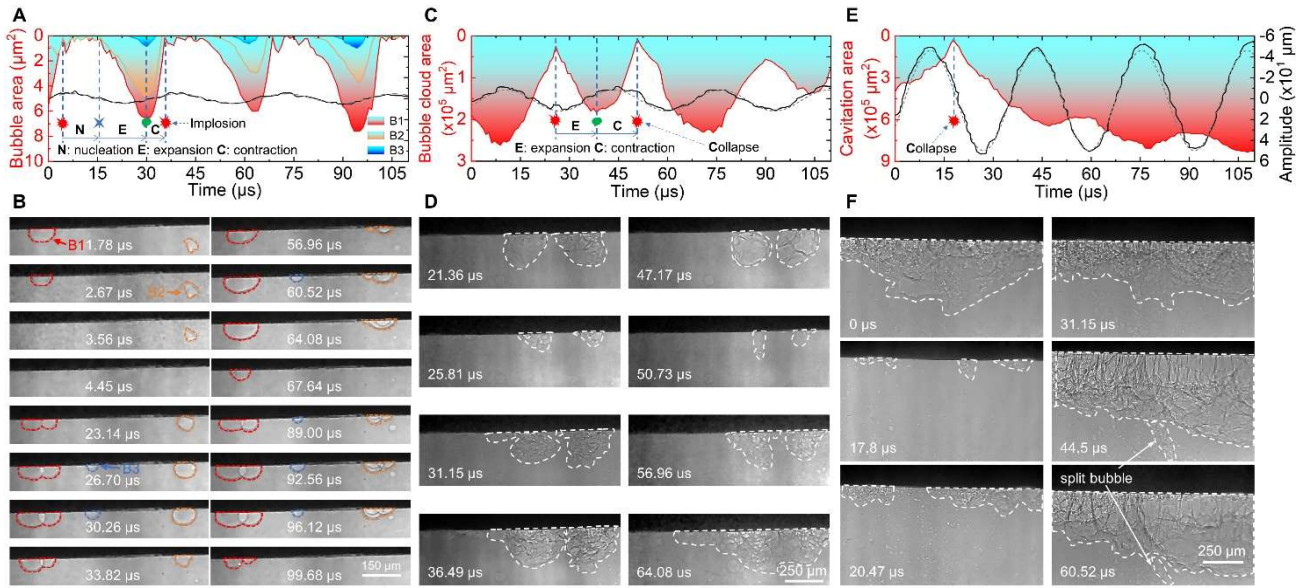

*Fig. 2. (A) Bubble area evolution of 3 typical bubbles (marked by B1, B2, and B3 in Fig. 2B) under 9 μm amplitude and (B) the corresponding images in 1 intra-train, showing the bubble nucleation, expansion, contraction and implosion dynamics with 886 ns time resolution. (C) Bubble cloud area evolution under 20 μm amplitude and (D) the corresponding images, showing the bubble cloud growth and coalescence dynamics. (E) Cavitation area evolution under 96 μm amplitude, and (F) the corresponding images, showing the cavitation bubble clouds collapse, reformation, growth, and coalescence. The grey dashed lines in (A), (C), and (E) are the fitted lines to the measured amplitudes. See the supplementary Videos 1, 2, and 3 for more vivid dynamic information.*

Fig. 2A showed that the sonotrode tip vibrated in a harmonic sine wave with the period (λ) of ~33.3 μs and 9 μm amplitude. Three typical bubbles were selected here. A single bubble (~100 μm diameter, pointed by the red arrow in Fig. 2B) pre-attached to the tip surface first imploded and rapidly split into two smaller bubbles when the tip started to vibrate upwards. The two bubbles then experienced expansion, contraction, implosion, and coalescence repeatedly in the following cycles. Bubble B2 (pointed by the orange arrow) first nucleated within the liquid ~100 μm below the tip, then oscillated with distortion and gradually moved towards the tip surface. After touching the tip surface, the bubble expanded further and imploded. Similar to bubble B1, it eventually split into smaller bubbles, acting as the bubble nuclei in the subsequent ultrasound cycles. The 3rd bubble B3 (30 μm diameter, indicated by the blue arrow) nucleated at the tip surface. Although the split was not observed due to its smaller size and the image spatial resolution, its growth and implosion were consistent with the behaviours of B1 and B2. The bubble area evolutions versus the applied amplitude were extracted from the images in Fig. 2B and shown in Fig. 2A for the 3 bubbles. Clearly, in any single ultrasound period (video 1 illustrates the critical points when bubble implosion and nucleation occurred), bubble nucleation occurred when the sonotrode started to move downwards. The bubble (area) continued to grow and reached its maximum size (the expansion period in Video 1). After the sonotrode started to move upwards, the bubble started the contraction period until its implosion when the sonotrode was near the highest position. More importantly, the annihilation of the bubble area (i.e., near zero area) corresponds to the moment of bubble implosion.

Fig. 2C shows that, as the tip vibration amplitude reached ~20 µm, the cavitation bubble area grew from several isolated ones to connected bubble clouds. After ~50-100 µs vibration, irregular cavitation bubble clouds were formed with their areas varying roughly in the form of sinusoid wave between 0.25 - 2.5 ×$10^5$ µm$^2$ and the similar period as the amplitude. These bubble clouds continued to grow, for example, at $t$ = 21.36 µs (Fig. 2D), there were two blurry bubble clouds at the tip surface, each experienced expansion, contraction, and collapse during the following acoustic cycle (see Video 2). One of the bubble clouds progressively approached the other and eventually coalesced into a larger cloud in the next acoustic cycle [31]. The images and video 2 vividly demonstrated how the bubble clouds grew, expanded, and coalesced to form a semi-connected cavitation area, which is the precursor for the full cavitation as described next. Fig. 2C shows that the rates of the bubble cloud expansion and collapse were nearly equal. Comparing with Fig. 2A, it is clear that the phase difference between the bubble cloud evolution and the tip vibration decreased. Again, the points of near zero bubble area correspond to the moment of bubble cloud collapsed and shock waves emission (the moments were marked clearly in Video 2).

When the tip vibration amplitude reached ~96 µm, Fig. 2E, 2F and video 3 show that several individual bubble clusters re-nucleated on the tip surface after a large bubble cloud collapsed. They grew, expanded, coalesced, and then formed a large bubble cloud again. Meanwhile, hundreds of microbubbles were formed instantly in the liquid below the tip and moved downwards. This phenomenon agreed with the reports [32, 33, 34] that dramatic pressure waves (i.e., shock waves) were emitted downward and hundreds of microbubbles formed when the primary cavitation bubble cloud collapsed. In the subsequent cycles, the bubble clouds continued to expand, oscillate and split bubble clouds into the liquid region below (as pointed by the white arrow in 2F). These split bubble clouds would expand, implode, and collapse further, emitting impulsive force into the liquid media[14, 35]. Such growth and coalescence of the bubble clouds were routinely observed in between 3 and 4 acoustic cycles before the bubble clouds in the cavitation zone stabilized and oscillated periodically. Once the full cavitation zone beneath the tip formed, the bubble clouds would oscillate for a few cycles before collapsing and emitting shockwave (pointed clearly in Video 3), and then embarking on another nucleation and growth cycle [34, 36]. Statistical analyses indicate that the average cavitation area and depth in steady state of bubble cloud oscillation was $7.5 \pm 0.6 \times 10^5$ µm$^2$ and $646 \pm 50$ µm respectively. Such information is critical for optimizing the ultrasonic process parameters (e.g., the power input and the working distance between the tip and sample) for the exfoliation of 2D functional materials.

*3.2 Shock waves enabled layer exfoliation of bulk graphite materials*

To further investigate the shock waves generated by cavitation bubble clouds implosion, the working distance between the sonotrode tip and HOPG sheet top surface was of utmost importance. The distance should be less than the sonotrode diameter (2 mm in this research) due to the exponential decay of the shock wave impact pressures with distance. It should also not be too close to avoid the shielding effect immediately below the sonotrode tip [13, 34]. Therefore, the working distance was set to ~1 mm and the top part of the HOPG sheet (~0.35 mm in height) appeared in the FOV as indicated in Fig. 1A(S2). Fig. 3B shows growth of the exfoliated layers of the HOPG sheet in DiWater for ~30 min (see the supplementary Video 4. 1 typical image from the 254 images in each train was selected to form this video). The graphite layers were gradually opened up under the cyclic impact of the arriving shock waves, forming roughly an arc-shaped region for the exfoliated layers.

We tracked and measured the deflection of 3 typical exfoliated layers as shown in Fig. 3B (marked by S1, S2, S3) versus the number of shock wave impacts (i.e., number of cavitation collapse). Fig. 3D shows clearly that the deflection followed approximately a parabolic curve. The layer in the centre (S3) produced the highest deflection, while that at the edge (S1) was the shortest. This is because the shock waves travelled slightly longer distance to reach S1, and the impact force decreased exponentially with the increase of travel distance. After ~10 mins (~ $6 \times 10^6$ shock wave impacts), the increase of the deflection slowed down and gradually approached to a steady-state plateau value. Fig. 3E shows the instantaneous deflection rate calculated from the curves in Fig. 3D.

To quantify the effect of one shock wave impact on layer exfoliation, Figs. 3F presents the relative deflection of the 3 graphite layers during 1 pulse train after 11 mins of processing.

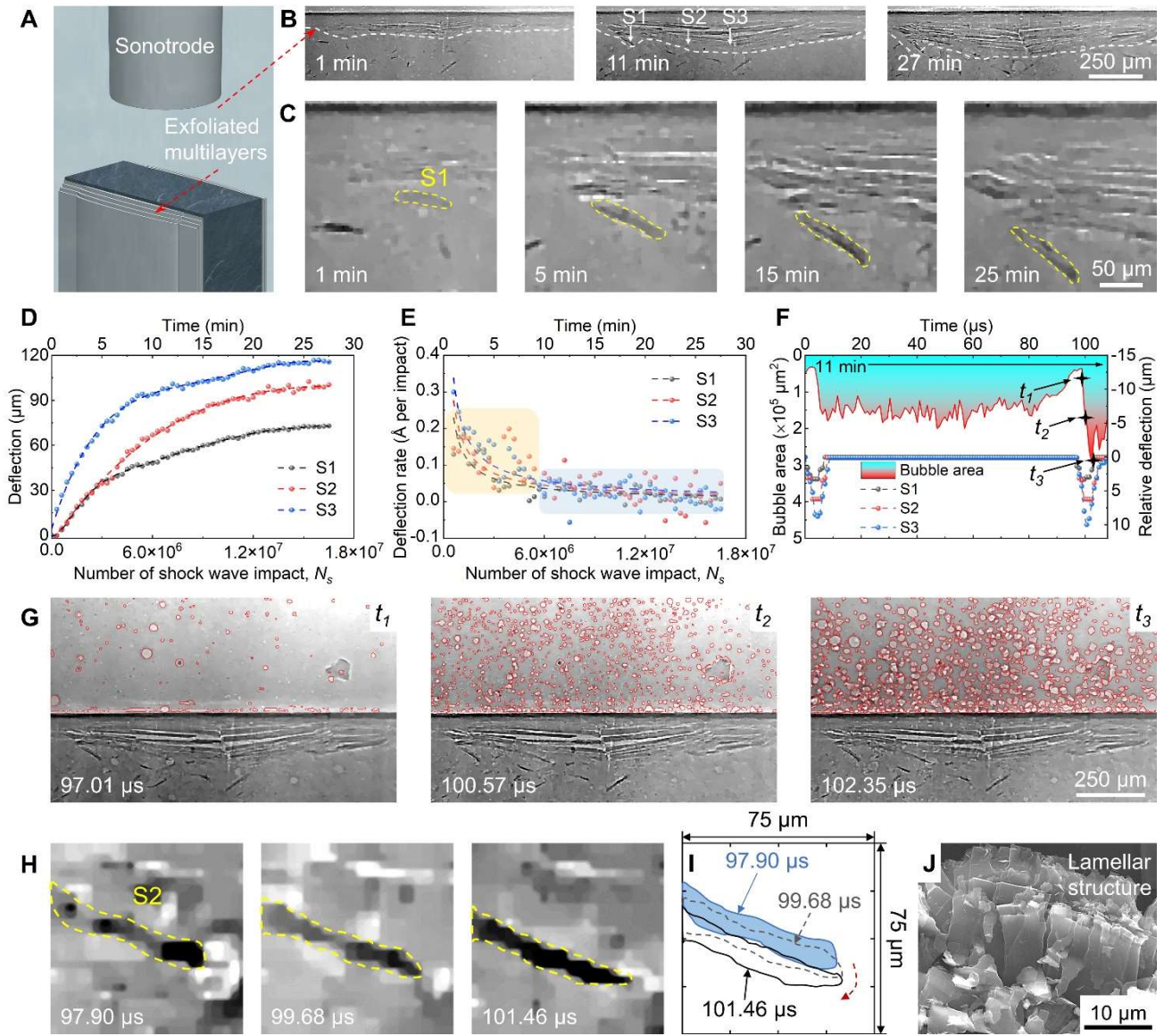

*Fig.3. (A) A schematic, showing the exfoliated graphite multilayers. (B) X-ray images, showing the layer exfoliation region in ~30 minutes. (C) Four X-ray images, highlighting the deflection of layer S1 at different time. (D and E) Deflection and instantaneous deflection rate of the 3 layers (pointed by the white arrows in Fig. 3B). (F) The relative deflection of the 3 graphite layers at the impact of two shock waves in 1 pulse train (The instantaneous near-zero bubble area indicated the arrival of the shock waves). (G)The X-ray images correspond to the instant of $t_1$, $t_2$ and $t_3$ in Fig. 3F, showing arrival of the shock waves and the resurge of tiny bubbles immediately afterwards. (H and I) The cropped and enlarged X-ray images at location S2 (Fig. 3B), showing the instantaneous movement of graphite layer S2 due to the shock waves. (J) The exfoliated graphite layer bundles consisting of many sub-µm lamellae after ULPE. (See Videos 4 and 5 for the exfoliation dynamics in min and sub-µs scale).*

Fig 3G shows the corresponding bubble behaviours. Clearly, at $t$ = 97.01 µs (a moment just before a detectable movement of the graphite layers), the bubble area in the FOV almost diminished (i.e., almost to zero). Within the next 5 µs, many hundreds of microbubbles were formed, accompanied by further expansion and downward movement (Fig. 3G ($t_2$) and ($t_3$)). This clearly illustrates the aftermath of the shock waves impact, transmitting through the DiWater and landing on top of the graphite surface, further splitting the graphite layers, as vividly shown in Fig. 3H. In the period ~5 µs, the graphite layers were deflected by 3 - 10 µm (Figs. 3F and 3I). The SEM image (Fig. 3J) shows that the exfoliated graphite layer bundles consisting of many sub-µm lamellae after ULPE. More importantly, the two shock waves were emitted at ~100 µs apart (Fig. 3F and 3G) in a subharmonic manner with a periodicity of 3λ [33, 34]. It is impossible to capture such shock wave-enabled graphite deflection/exfoliation by conventional optical imaging techniques at such combined temporal and spatial resolution. Based on this, the number of shock waves ($N_s$) landed onto the graphite layers can be calculated, which can be used to derive the exfoliation rate for the HOPG, i.e., ~0.15 Å per impact in the early stage (see the light yellow region in Fig. 3E), and ~0.015 Å per impact in the steady-state stage (see the light blue area in Fig. 3E). Clearly the 10 mins in the early stage of ultrasound processing is very effective and efficient for layer exfoliation.

*3.3 Fatigue effects of the cyclic action of shock waves on the layer exfoliation dynamics*

Fig. 4, Videos 6 and 7 showed the exfoliation dynamics of a thin GF in DiWater for 210 s. Because many more structural defects existed in the GF compared to those in the bulk HOPG sheet, the GF was exfoliated much more easily during the ULPE process. Therefore, the sonotrode tip was placed ~1 mm above the upper edge of the FOV (i.e., a distance of ~1.3 mm between the GF and the tip) and the ultrasound was set to run for 2 s in every 13 s, i.e., synchronised with the recording and data transfer of the camera in every 15 s to capture the longer time scale exfoliation dynamics, hence a total processing time of ~210 s was applied. Fig. 4A shows that the GF was gradually exfoliated into seaweed-like branches (i.e., layer bundles). Eventually, the rightmost branch collapsed and out of the view field in 210 s. Figs. 4B and 4C present the exfoliated length and angle of the rightmost branch in Fig. 4A (see the 150 s image) versus ultrasound processing time. Fig. 4F shows the corresponding polar contour plot. Fig. 4B shows that exfoliation of the GF has four stages: (I) before 50 s, there was an incubation period where the flake remained relatively stable, and only a few layer separation events occurred, (II) in 50 - 100 s, separation and exfoliation of the layers accelerated with many new layer bundles formed and the exfoliation length quickly

reached to ~350 µm. These layer bundles oscillated in ~10 degrees range, (III) in 100 – 150 s, the split layer bundles were deflected to wider angle ranges (upto ~30 degrees), but the increase of the layer length slowed down, (IV) after 150 s, as the deflection angle further widened, the layer exfoliation accelerated again, quickly reached ~500 µm.

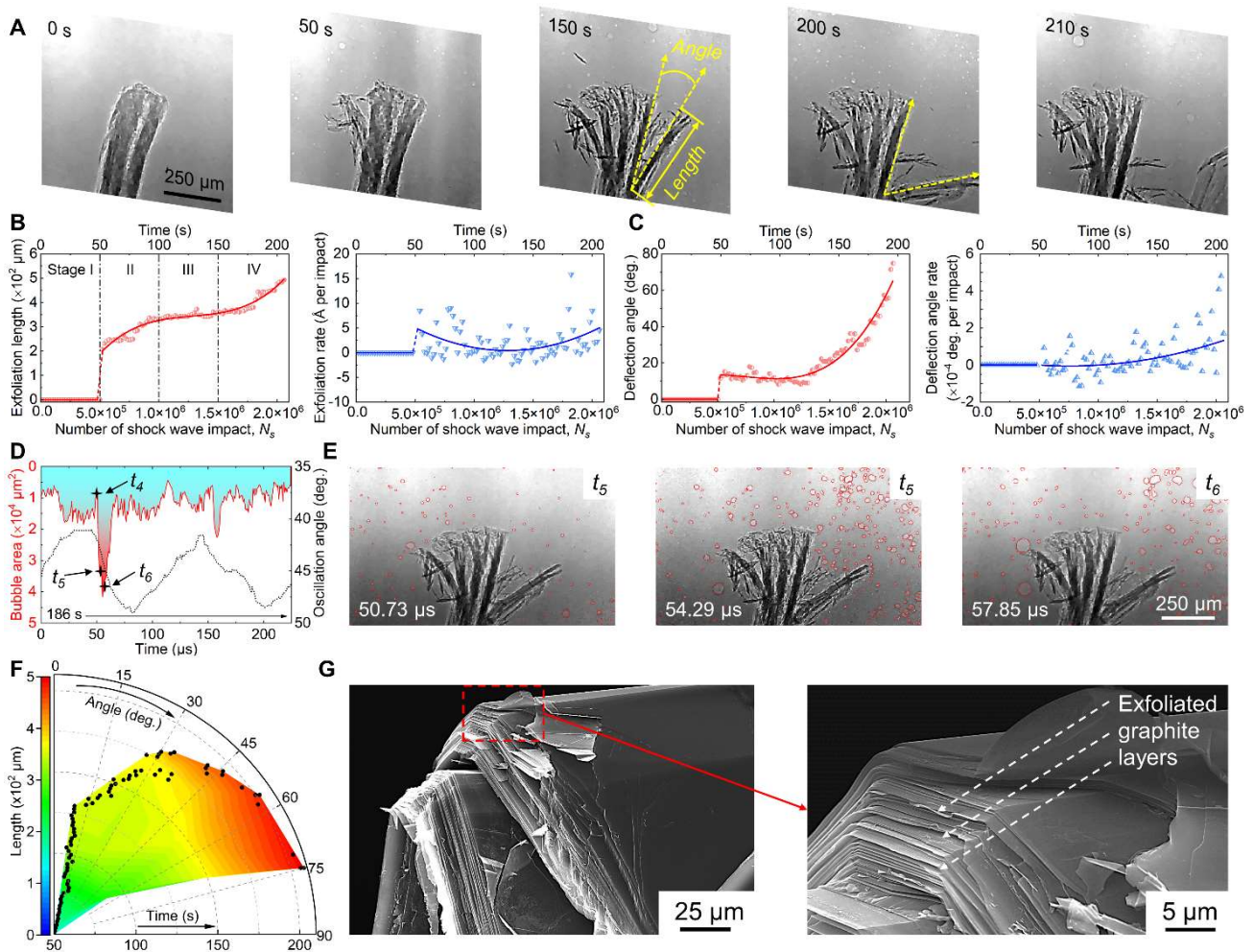

*Fig. 4. (A) A series of typical X-ray images, showing the layer exfoliation dynamics of a thin GF in 210 s. (B) Exfoliation length and instantaneous exfoliation rate (the rightmost graphite branch in Fig. 4A). (C) Deflection angle and deflection rate versus the number of shock wave impacts. (D) The oscillating angles and the corresponding bubble area evolution in 1 pulse train. (E) The X-ray images corresponding to the instant of $t_4$, $t_5$ and $t_6$ in Fig. 4D, showing the arrival of the shock wave and impact onto the graphite branches. (F) The polar contour map, showing the evolution of exfoliation length and angle versus time. The black dots in the plot represent the position of the branch at different times. (G) SEM images of the post-processed GF, showing the exfoliated multiple graphite layers. (See the supplementary Videos 6 and 7 for the vivid dynamic information in s and sub-µs scale).*

Again, we studied the exfoliation dynamics in one pulse train (i.e., with 886 ns time resolution), Figs. 4D and 4E show one case after ~186 s of ultrasound processing. Clearly, at the moment when the graphite branch started to oscillate, hundreds of microbubbles were formed instantly in the liquid (aftermath of the shock waves, see Fig. 4E and Video 7). In addition, the oscillation and bubble area evolution (see Fig. 4D) followed a subharmonic periodicity of ~3λ. This agreed well with the observations in Fig. 3F. Again, it confirmed that the layer deflection and exfoliation were due to the arrival of the shock waves generated by the subharmonic collapses of the cavitation bubble clouds [13]. Fig. 4G showed that the exfoliated layer bundles consist of hundreds of sub-μm size and kink multilayers. Such graphite layer bundles would be further exfoliated in the liquid with sufficient ultrasonic processing time to achieve the production of graphite nanoflakes or multilayer graphene [13, 20, 37, 38].

A simple calculation indicated that, in the 210 s ultrasound processing, the GF was subjected to a total of ~2.1×$10^6$ shock wave impacts. Based on this, the instantaneous exfoliation and deflection angle rate under a single shock wave impact were calculated. Fig. 4B shows that, after the incubation period, the exfoliation rate was as high as ~5 Å per shock wave impact, then it gradually decreased to ~0.5 Å per impact. Compared to the exfoliation dynamics of the HOPG sheet, The GF showed a similar steady-state exfoliation stage, where the exfoliated length hardly changed with the exfoliated rate of ~0.5 Å per impact. However, in the GF case, the deflection angle increased to or above ~30 degrees. The exfoliation rate and deflection angle rate accelerated again, reaching ~5 Å and ~1.5×$10^{-4}$ degree per impact in the final stage.

McAllister et al. [7] calculated that the pressure required to overcome the van der Waals bond force between two graphite multilayer sheets was ~2.5 MPa. While the peak acoustic pressure beneath the sonotrode tip was ~1.5 MPa, and the pressure reached the GF in our case would be ~1 MPa in the DiWater [16]. Although the impact pressure by a single shock wave was not able to separate the well-bonded graphite layers with one impact, for those weak layers with sufficient defects such as the graphite flakes in this study, clearly they are exfoliated progressively by the cyclic fatigue action of the shock wave impact [13] in a time scale that is meaningful to industry. In this work, we only used DiWater as the liquid medium to transmit the ultrasonic waves, hence any chemical effects due to the additions of other solvents on the exfoliation dynamics were excluded here. Our systematic MHz imaging observation and the consistent results from two graphite materials revealed unambiguously that the fatigue exfoliation mechanism due to the cyclic shock wave impacts produced by ultrasound cavitation is the most important and dominant mechanism for ULPE of graphite materials. These new findings will have great impact for developing industrial upscaling strategies for ultrasonic exfoliation of 2D materials in the near future.

## 4. Conclusions

Using the XFEL MHz imaging with ~25 fs X-ray pulses, we have visualised and revealed the ultrasound liquid phase exfoliation dynamics with 886 ns time resolution and 3.2 μm spatial resolution, the combined resolutions that have never been achieved before. By taking the advantage of the unique X-ray pulse train time structure synchronised with two full field Shimadzu cameras, we have tracked and quantified the exfoliated layer growth dynamics up to 500 μm in layer length and thousands of seconds in operando conditions that are directly relevant to industrial practice. The important findings of this research are:

1) The amplitude-controlled ultrasound cavitation bubbles nucleation, growth and implosion dynamics are revealed and quantified systematically. At an amplitude of less than ~20 μm, individual bubble and semi-connected bubble cloud formed. The bubble expansion and contraction follow the harmonic sine wave according to the ultrasound applied. At an amplitude of 96 μm, a full cloud cavitation zone was developed, and the cloud area became continuous and relatively stable. Bubble cloud reformation, growth, and coalescence were all quantified, and they occur periodically in 3 acoustic cycles.
2) Bubble cloud implosion companied by shock wave emission occurred with a subharmonic periodicity of 3λ, deflecting and exfoliating graphite layers cyclically in a time scale of a few μs through the fatigue mechanism. For the bulk HOPG sheet, the deflection rate was ~0.15 Å per shock wave impact in the early stage, slowed down to ~0.015 Å per impact in the steady-state stage. For the graphite flakes, the exfoliation rate was as high as ~5 Å per shock wave impact in the early and later stage, while maintained at ~0.5 Å per impact in the steady-state period.

## 5. Acknowledgment

The authors would acknowledge the European XFEL in Schenefeld, Germany, for provision of the X-ray free-electron laser beamtime (Proposal No. 3100) at SPB/SFX SASE1, the EPSRC funding from the UK Engineering and Physical Sciences Research Council (Grant No. EP/R031819/, EP/R031665/1, EP/R031401/1, EP/R031975/1) and the Internal EuXFEL R&D project "MHz microscopy at EuXFEL: From demonstration to method", 2020 – 2022,

HORIZON-EIC-2021-PATHFINDEROPEN-01-01, MHz--TOMOSCOPY project, Grant agreement: 101046448. Xiang and Huang would also like to acknowledge the PhD Studentships awarded jointly by the Hull University and China Scholarship Council scheme. Patrik Vagovič acknowledges the technical support of Thomas Dietze.

## 6. Author Contributions

Experimental design and execution (Kang Xiang, Shi Huang, Hongyuan Song, Jiawei Mi, and Patrik Vagovič); Conceiving and establishment of MHz imaging method and apparatus at SPB/SFX instrument (Patrik Vagovič) , measurements (Kang Xiang, Shi Huang, Jiawei Mi, Valerio Bellucci, Sarlota Birnšteinová, Raphael de Wijn, Jayanath C. P. Koliyadu, Faisal Koua, Adam Round, Tokushi Sato, Marcin Sikorski, Yuhe Zhang, Eleni Myrto Asimakopoulou, Pablo Villanueva-Perez, Richard Bean, Patrik Vagovič); data analysis (Kang Xiang, Sarlota Birnšteinová, Yuhe Zhang, Pablo Villanueva-Perez, Jiawei Mi), sample preparation (Kang Xiang, Shi Huang, Hongyuan Song, Ekaterina Round, V. Bazhenov, Abhisakh Sarma, Jiawei Mi); Funding acquisition (Kyriakos Porfyrakis, Iakovos Tzanakis, Dmitry G. Eskin, Nicole Grobert, and Jiawei Mi); Kang Xiang and Jiawei Mi wrote the manuscript with inputs from all other authors.

## 7. Competing interests

The authors declare that there are no competing interests.

## 8. References

1. Geim AK. Graphene: status and prospects. *science* 2009, **324**(5934)**:** 1530-1534.

2. Khan K, Tareen AK, Aslam M, Wang R, Zhang Y, Mahmood A*, et al.* Recent developments in emerging two-dimensional materials and their applications. *Journal of Materials Chemistry C* 2020, **8**(2)**:** 387-440.

3. Li X, Zhi L. Graphene hybridization for energy storage applications. *Chemical Society Reviews* 2018, **47**(9)**:** 3189-3216.

4. Novoselov KS, Fal′ko VI, Colombo L, Gellert PR, Schwab MG, Kim K. A roadmap for graphene. *nature* 2012, **490**(7419)**:** 192-200.

5. Cravotto G, Cintas P. Sonication-assisted fabrication and post-synthetic modifications of graphene-like materials. *Chemistry* 2010, **16**(18)**:** 5246-5259.

6.  Niu L, Coleman JN, Zhang H, Shin H, Chhowalla M, Zheng Z. Production of two-dimensional nanomaterials via liquid-based direct exfoliation. *Small* 2016, **12**(3): 272-293.

7.  McAllister MJ, Li J-L, Adamson DH, Schniepp HC, Abdala AA, Liu J*, et al.* Single sheet functionalized graphene by oxidation and thermal expansion of graphite. *Chemistry of materials* 2007, **19**(18): 4396-4404.

8.  Xu Y, Cao H, Xue Y, Li B, Cai W. Liquid-Phase Exfoliation of Graphene: An Overview on Exfoliation Media, Techniques, and Challenges. *Nanomaterials (Basel)* 2018, **8**(11).

9.  Parviz D, Irin F, Shah SA, Das S, Sweeney CB, Green MJ. Challenges in Liquid-Phase Exfoliation, Processing, and Assembly of Pristine Graphene. *Adv Mater* 2016, **28**(40): 8796-8818.

10. Coleman JN, Lotya M, O'Neill A, Bergin SD, King PJ, Khan U*, et al.* Two-dimensional nanosheets produced by liquid exfoliation of layered materials. *Science* 2011, **331**(6017): 568-571.

11. Nicolosi V, Chhowalla M, Kanatzidis MG, Strano MS, Coleman JN. Liquid exfoliation of layered materials. *Science* 2013, **340**(6139): 1226419.

12. Hernandez Y, Nicolosi V, Lotya M, Blighe FM, Sun Z, De S*, et al.* High-yield production of graphene by liquid-phase exfoliation of graphite. *Nature nanotechnology* 2008, **3**(9): 563-568.

13. Morton JA, Khavari M, Qin L, Maciejewska BM, Tyurnina AV, Grobert N*, et al.* New insights into sono-exfoliation mechanisms of graphite: In situ high-speed imaging studies and acoustic measurements. *Materials Today* 2021.

14. Qin L, Maciejewska BM, Subroto T, Morton JA, Porfyrakis K, Tzanakis I*, et al.* Ultrafast synchrotron X-ray imaging and multiphysics modelling of liquid phase fatigue exfoliation of graphite under ultrasound. *Carbon* 2022, **186:** 227-237.

15. Priyadarshi A, Khavari M, Subroto T, Conte M, Prentice P, Pericleous K*, et al.* On the governing fragmentation mechanism of primary intermetallics by induced cavitation. *Ultrason Sonochem* 2021, **70:** 105260.

16. Qin L, Porfyrakis K, Tzanakis I, Grobert N, Eskin DG, Fezzaa K*, et al.* Multiscale interactions of liquid, bubbles and solid phases in ultrasonic fields revealed by multiphysics modelling and ultrafast X-ray imaging. *Ultrason Sonochem* 2022, **89:** 106158.

17. Kaur A, Morton JA, Tyurnina AV, Priyadarshi A, Holland A, Mi J*, et al.* Temperature as a key parameter for graphene sono-exfoliation in water. *Ultrason Sonochem* 2022, **90:** 106187.

18. Tyurnina AV, Tzanakis I, Morton J, Mi J, Porfyrakis K, Maciejewska BM, *et al.* Ultrasonic exfoliation of graphene in water: A key parameter study. *Carbon* 2020, **168:** 737-747.

19. Li X, Wang X, Zhang L, Lee S, Dai H. Chemically derived, ultrasmooth graphene nanoribbon semiconductors. *science* 2008, **319**(5867)**:** 1229-1232.

20. Alaferdov AV, Gholamipour-Shirazi A, Canesqui MA, Danilov YA, Moshkalev SA. Size-controlled synthesis of graphite nanoflakes and multi-layer graphene by liquid phase exfoliation of natural graphite. *Carbon* 2014, **69:** 525-535.

21. Lotya M, King PJ, Khan U, De S, Coleman JN. High-concentration, surfactant-stabilized graphene dispersions. *ACS nano* 2010, **4**(6)**:** 3155-3162.

22. Wang B, Tan D, Lee TL, Khong JC, Wang F, Eskin D, *et al.* Ultrafast synchrotron X-ray imaging studies of microstructure fragmentation in solidification under ultrasound. *Acta Materialia* 2018, **144:** 505-515.

23. Tyurnina AV, Morton JA, Subroto T, Khavari M, Maciejewska B, Mi J, *et al.* Environment friendly dual-frequency ultrasonic exfoliation of few-layer graphene. *Carbon* 2021, **185:** 536-545.

24. Vagovič P, Sato T, Mikeš L, Mills G, Graceffa R, Mattsson F, *et al.* Megahertz x-ray microscopy at x-ray free-electron laser and synchrotron sources. *Optica* 2019, **6**(9).

25. Sobolev E, Zolotarev S, Giewekemeyer K, Bielecki J, Okamoto K, Reddy HKN, *et al.* Megahertz single-particle imaging at the European XFEL. *Communications Physics* 2020, **3**(1).

26. Sato T, Letrun R, Kirkwood HJ, Liu J, Vagovič P, Mills G, *et al.* Femtosecond timing synchronization at megahertz repetition rates for an x-ray free-electron laser. *Optica* 2020, **7**(6).

27. Vakili M, Bielecki J, Knoska J, Otte F, Han H, Kloos M, *et al.* 3D printed devices and infrastructure for liquid sample delivery at the European XFEL. *J Synchrotron Radiat* 2022, **29**(Pt 2)**:** 331-346.

28. Van Nieuwenhove V, De Beenhouwer J, De Carlo F, Mancini L, Marone F, Sijbers J. Dynamic intensity normalization using eigen flat fields in X-ray imaging. *Opt Express* 2015, **23**(21)**:** 27975-27989.

29. Birnsteinova S, de Lima DEF, Sobolev E, Kirkwood HJ, Bellucci V, Bean RJ, *et al.* Online dynamic flat-field correction for MHz Microscopy data at European XFEL. *arXiv preprint arXiv:230318043* 2023.

30. Buakor K, Zhang Y, Birnšteinová Š, Bellucci V, Sato T, Kirkwood H, *et al.* Shot-to-shot flat-field correction at X-ray free-electron lasers. *Optics Express* 2022, **30**(7)**:** 10633-10644.

31. Postema M, Marmottant P, Lancee CT, Hilgenfeldt S, de Jong N. Ultrasound-induced microbubble coalescence. *Ultrasound Med Biol* 2004, **30**(10): 1337-1344.

32. Birkin PR, Offin DG, Vian CJ, Leighton TG. Multiple observations of cavitation cluster dynamics close to an ultrasonic horn tip. *J Acoust Soc Am* 2011, **130**(5): 3379-3388.

33. Yusuf L, Symes MD, Prentice P. Characterising the cavitation activity generated by an ultrasonic horn at varying tip-vibration amplitudes. *Ultrason Sonochem* 2021, **70:** 105273.

34. Khavari M, Priyadarshi A, Morton J, Porfyrakis K, Pericleous K, Eskin D*, et al.* Cavitation-induced shock wave behaviour in different liquids. *Ultrason Sonochem* 2023, **94:** 106328.

35. Shchukin DG, Skorb E, Belova V, Mohwald H. Ultrasonic cavitation at solid surfaces. *Adv Mater* 2011, **23**(17): 1922-1934.

36. Khavari M, Priyadarshi A, Hurrell A, Pericleous K, Eskin D, Tzanakis I. Characterization of shock waves in power ultrasound. *Journal of Fluid Mechanics* 2021, **915**.

37. Li Z, Young RJ, Backes C, Zhao W, Zhang X, Zhukov AA*, et al.* Mechanisms of Liquid-Phase Exfoliation for the Production of Graphene. *ACS Nano* 2020, **14**(9): 10976-10985.

38. Turner P, Hodnett M, Dorey R, Carey JD. Controlled Sonication as a Route to in-situ Graphene Flake Size Control. *Sci Rep* 2019, **9**(1): 8710.